\documentclass{article}
\usepackage{amsmath, amsfonts,amssymb}
\usepackage{epsfig}
\usepackage{cite, bbm, color}

\newcommand{\eq}[1]{\begin{equation}\label{#1}}
\newcommand{\en}{\end{equation}}
\newcommand{\ba}{\begin{eqnarray}}
\newcommand{\ea}{\end{eqnarray}}
\newcommand{\rmi}[1]{{\mbox{\scriptsize #1}}}

\newcommand{\ear}[1]{\begin{eqnarray}\label{#1}}
\newcommand{\enar}{\end{eqnarray}}
\newcommand{\re}{{\rm{Re}}}
\newcommand{\dd}{{\rm{d}}}
\newcommand{\tr}{{\rm Tr\,}}

\renewcommand{\(}{\left(}
\renewcommand{\)}{\right)}

\def\openone{\rlap 1\kern 0.22ex 1}

\newcommand{\bi}{\begin{itemize}}
\newcommand{\ei}{\end{itemize}}

\begin{document}

\begin{titlepage}
\begin{flushright} CERN-PH-TH/2009-191 \end{flushright}
\vspace*{0.5cm}
\begin{center}
{\Large\bf The phase diagram of Yang-Mills theory \\
with a compact extra dimension}
\end{center}
\vskip1.3cm
\centerline{Philippe~de~Forcrand,$^{a,b}$ Aleksi~Kurkela$^{a}$ and  Marco~Panero$^{a}$}
\vskip1.5cm
\centerline{\sl  $^a$ Institute for Theoretical Physics,  ETH Zurich, CH-8093 Zurich, Switzerland}
\vskip0.5cm
\centerline{\sl  $^b$ CERN, Physics Department, TH Unit, CH-1211 Geneva 23, Switzerland}
\vskip0.5cm
\begin{center}
{\sl  e-mail:} \hskip 6mm \texttt{forcrand@phys.ethz.ch, kurkela@phys.ethz.ch, panero@phys.ethz.ch}
\end{center}
\vskip1.0cm
\begin{abstract}
We present a non-perturbative study of the phase diagram of SU(2) Yang-Mills theory in a five-dimensional spacetime with a compact extra dimension. 
The non-renormalizable theory is regularized on an anisotropic lattice and investigated through numerical simulations in a regime characterized by a hierarchy between the scale of low-energy physics, the inverse compactification radius, and the cutoff scale. We map out the structure of the phase diagram and the pattern of lines corresponding to fixed values of the ratio between the mass of the fifth component of the gauge field and the non-perturbative mass gap of the four-dimensional modes. We discuss different limits of the model, and comment on the implications of our findings.
\end{abstract}
\end{titlepage}

\section{Introduction}
\label{sec:intro}

Gauge theories defined in a spacetime with compact extra dimensions offer a very interesting platform for model building and phenomenology beyond the Standard Model, and provide appealing theoretical scenarios for Grand Unification and electroweak symmetry breaking, for the fermion hierarchy problem, and for the strong CP problem~\cite{Scherk:1979zr, Manton:1979kb, Fairlie:1979at, Hosotani:1983xw, Khlebnikov:1987zg, Dienes:1998vg, ArkaniHamed:1998rs, ArkaniHamed:1998nn, Dienes:1999gw, Chaichian:2001nx, Scrucca:2003ra, Csaki:2003dt, Khlebnikov:2004am, Bezrukov:2008da}. 

Although a gauge theory in $d>4$ is generally non-renormalizable, and thus cannot be considered as a fundamental theory, it can be interpreted as a low-energy effective description of an underlying theory. The effective theory is consequently a description valid only for energies lower than a cutoff $\Lambda$, which is the scale where the microscopic details of the underlying theory become relevant and the effective description breaks down.
In particular, for a compact extra dimension of size $L_5$, this effective theory description is expected to be sensible only if there is at least a modest scale separation between the inverse cutoff $\Lambda^{-1}$ and the compactification length, $L_5 \Lambda \gg 1$. 

Assuming that the relevant correlation lengths are much longer than the inverse cutoff $\Lambda^{-1}$, the actual ultra-violet completion of the theory can be traded for any simpler regularization, introducing only finite systematic errors. In particular, choosing the lattice regularization, one can obtain robust non-perturbative predictions for the low-energy sector of the effective theory from numerical simulations. In this work, we investigate a simple model of this type: non-Abelian SU(2) Yang-Mills (YM) theory in a flat (4+1)d Euclidean spacetime. 

The properties of 5d YM models on the lattice have been studied in several works, including Refs.~\cite{Creutz:1979dw, Kawai:1992um, Nishimura:1996pe, Beard:1997ic, Chandrasekharan:1996ih, Brower:1997ha, Ejiri:2000fc,Ejiri:2002ww, Dimopoulos:2000ej, Dimopoulos:2000iq, Dimopoulos:2001un, Farakos:2002zb, Irges:2004gy, Irges:2006zf, Irges:2006hg, Irges:2007qq, Irges:2009bi, Irges:2009qp, Hosotani:2007kn, DelDebbio:2008hb, Sakamoto:2009hb, Ishiyama:2009bk}. 
In fact, our approach builds on the evidence from previous numerical studies, showing that SU(2) and SU(3) YM theories defined on an isotropic 5d space exhibit a Coulomb phase in the weak (bare) coupling regime~\cite{Creutz:1979dw, Kawai:1992um, Nishimura:1996pe, Beard:1997ic}, and that upon compactification of one of the directions, the theory undergoes dimensional reduction for any compactification radius~\cite{Chandrasekharan:1996ih, Brower:1997ha}.  It is the existence of the Coulomb phase and dimensional reduction that guarantees that the relevant correlation lengths are large compared to the cutoff and thus justifies the use of the lattice regularization.

In our model, the dimensional reduction allows one to describe the low-energy sector of the (4+1)d theory using 4d YM theory, which has a non-perturbatively generated string tension proportional to
\eq{mass_gap}
\sigma \propto \exp \left( - \frac{24 \pi^2 L_5}{11g_5^2} \right),
\en
where $g_5$ denotes the gauge coupling of the (4+1)d theory. Measuring all quantities in units of the string tension (instead of the cutoff) allows one to build a consistent definition of the lattice spacing and of lines of constant low-energy physics, in terms of 4d quantities. 

This paper is structured as follows: after defining the lattice formulation of the model in Section~\ref{sec:theory}, we study the phase diagram of the theory in Sec.~\ref{sec:phase_structure}, where we also discuss some qualitative analogies of the model with compact U(1) lattice gauge theory in (3+1)d. Our investigation of the properties of the dimensionally reduced phase is reported in Sec.~\ref{sec:dim_red}, while in Sec.~\ref{sec:contlim} we comment on possible strategies of removing the cutoff and why they must eventually fail. Finally, we summarize our findings and present an outlook on future research directions in Sec.~\ref{sec:conclusions}. Some technical details of our calculation are discussed in the Appendix.

\section{The model}
\label{sec:theory}
As a prototype for an extra-dimensional gauge model, we study SU(2) Yang-Mills theory in (4+1)d Euclidean spacetime, where one of the dimensions is taken to be periodic with a finite extent $L_5$ and the gauge fields are taken to obey periodic boundary conditions along the compact direction. 
The model is formally defined by the path integral
\begin{equation}
Z = \int\mathcal{D}A \textrm{e}^{-S_E}, \quad \quad S_E = \int {\dd}^4x  \int_0^{L_5}{\dd} x_5 \; \frac{1}{2 g_5^2 }\tr F_{MN}^2,\label{CS}
\end{equation}
where the $M, N$ indices run from 1 to 5. Since the bare coupling $g_5^2$ has the dimensions of a length, this model is non-renormalizable, and the continuum action in Eq.~(\ref{CS}) defines it only up to a regularization, i.e. as a theory with a cutoff $\Lambda$. The model is thus parametrized by two dimensionless ratios, namely $L_5\Lambda$ and $g^2_5 \Lambda$. 

To study the model at non-zero coupling values, we regularize it on an anisotropic lattice with a lattice spacing $a$ in the four usual directions, letting the lattice spacing $a_5$ in the fifth direction take an independent value. Furthermore, let $N_5$ denote the extent of the compact direction in units of $a_5$, such that $L_5=a_5N_5$. We define the model on the  lattice through the Wilson action
\eq{eq:lattice_action}
S_E^L=\frac{\beta_5}{\gamma} \sum_{x,1 \le M < N \le 4} \left[1 - \frac{1}{2} \re\tr P_{MN}(x)\right]+\gamma \beta_5 \sum_{x,1 \le M \le 4} \left[ 1 - \frac{1}{2} \re\tr P_{M5}(x)\right], 
\en
with
\eq{eq:beta5}
\beta_5=\frac{4 a}{g_5^2}.
\en
In the weak coupling limit, the anisotropy factor $\gamma$ reduces to the ratio of the lattice spacings
\begin{equation}
\lim_{\beta_5\rightarrow \infty} \frac{a}{a_5}=\gamma.
\end{equation}

The non-renormalizability of the model implies that the limit in which both $a$ and $a_5$ are taken to zero leads to divergences which cannot be absorbed in coefficients of an action with a finite number of terms, and thus the cutoff cannot be removed. However, it is possible to take the lattice spacing in the compact direction to zero, $a_5\rightarrow 0$, while keeping $N_5/\gamma\equiv \widetilde{N}_5\approx L_5/a$ and the lattice spacing $a$ in the other four directions fixed.\footnote{This leads to a quantum-mechanical system defined on a 4d lattice with finite spacing $a$, and is not in conflict with non-renormalizability issues.} 
In this work, we present results obtained from extrapolations to this limit, which offers two advantages:
\begin{itemize}
\item it reduces the number of parameters characterizing the model from three ($N_5$, $\gamma$, and $\beta_5$) down to only two ($\widetilde{N}_5$ and $\beta_5$) like in the continuum theory; 
\item it suppresses cutoff effects in systems where the $L_5/a$ ratio is not very large. Practically, this is very important in our numerical calculations, since the 4d correlation length of the system strongly depends on this ratio.
\end{itemize}

After taking the continuum limit in the fifth dimension, the 4d lattice spacing $a$ cannot be taken to zero, but one can still approach the continuum limit of the theory by tuning the correlation length of the lattice model $\xi$ to be much larger than the lattice spacing $a$: in such a limit, the low-energy sector of the model becomes insensitive to the lattice discretization artifacts, and the continuum 4d symmetries are  restored. 

We have investigated the phase diagram and various correlation lengths of the model non-perturbatively via Monte~Carlo simulations, calculating two-point correlation functions of Polyakov loops along the extra dimension, as well as along the four ordinary directions (torelons). In particular, the torelon mass $E_t$, the string tension $\sigma$, and the energy of  the first Kaluza-Klein excitation $E_{KK}$ can be extracted from the two-point correlation function of zero transverse-momentum torelons $\tilde{P}_x(r)$ along one of the space-like directions of size $L$: for asymptotically large separations $\tau$ one has 
\eq{Kaluza_Klein}
\langle \tilde{P}_x^\dagger(r) \tilde{P}_x(0)\rangle \simeq A_1 e^{-E_t r}+A_2 e^{-E_{KK} r},
\en
with 
\eq{Kaluza_Klein_masses}
E_t = \sigma L - \frac{\pi}{L}\frac{d_t}{6},\;\;\;\;\; E_{KK} = \sqrt{E_t^2 + \(\frac{2\pi}{L_5}\)^2}.
\en
Note that, besides the term proportional to the string tension, $E_t$ also includes a L\"uscher-term correction~\cite{Luscher:1980fr, Luscher:1980ac}, which is the leading contribution coming from the massless fluctuations of the torelon along the $d_t=2$ transverse dimensions (see also Ref.~\cite{deForcrand:1984cz} for a discussion).

\section{Phase diagram}
\label{sec:phase_structure}

The phase structure of 5d SU(2) Yang-Mills theory was first studied non-perturbatively in Ref.~\cite{Creutz:1979dw} for an isotropic action on a lattice with same extents in all directions. Unless a dimensionful scale $L_5$ is introduced by compactification, the theory has only one parameter, namely the lattice coupling $\beta_5$. In this case, the phase diagram consists of a confining phase in the strong-coupling regime at small $\beta_5$ values, and a Coulomb-like phase at weak coupling, i.e. at large $\beta_5$. 
The confining phase is separated from the Coulomb phase by a transition, which (based on numerical evidence from simulations on a $4^5$ lattice) Ref.~\cite{Creutz:1979dw} located at $\beta_5^c=1.65$. This is identified as a first-order bulk phase transition, revealed by a discontinuity in the plaquette expectation value and not related to any breaking of a global symmetry. As the confining phase is not analytically connected to the weak-coupling regime, we consider it merely as a lattice artifact, having little to do with the continuum theory.

Since in the absence of a compactification the weak-coupling phase is characterized by an infinite correlation length, we expect that the compactification of the fifth dimension leads to spontaneous breaking of the center symmetry and to dimensional reduction, \emph{for any finite compactification radius}. For this reason we denote this phase, characterized by a non-zero Polyakov loop expectation value ($\langle P_5 \rangle\neq 0$), as the ``dimensionally reduced phase''. Our numerical results indeed confirm the existence of two phases also for the anisotropic model defined by Eq.~(\ref{eq:lattice_action}), for all compactification lengths $\widetilde{N}_5$.  

\begin{figure}
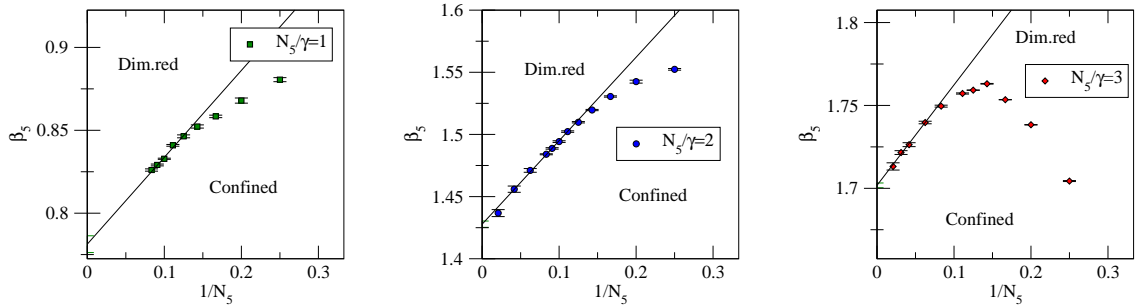
  
\centerline{\includegraphics*[width=0.3\textwidth]{fig1a.eps}\hfill\includegraphics*[width=0.3\textwidth]{fig1b.eps}\hfill\includegraphics*[width=0.3\textwidth]{fig1c.eps}}
\caption{The effect of the discretization due to finite $N_5$ on the pseudo-critical coupling $\beta_c(N_5/\gamma)$. The phase transition breaks the center symmetry of the Polyakov loop along the extra dimension, such that $\langle P_5 \rangle$ vanishes in the confining phase but has a non-zero expectation value in the dimensionally reduced phase. The pseudo-critical coupling is defined through the maximum of the susceptibility $\chi_5=\langle |P_5|^2 \rangle -\langle |P_5| \rangle^2$.}
\label{finite_N5_effects_on_betac}
\end{figure}

\begin{figure}
\begin{center}
\includegraphics*[width=0.5\textwidth]{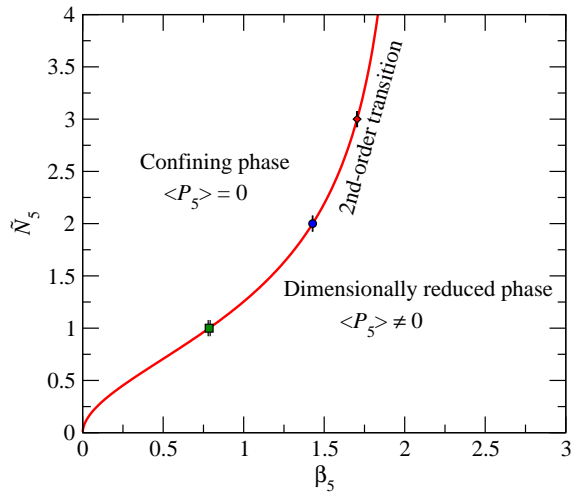}
\end{center}
\caption{The phase diagram of 5d SU(2) lattice gauge theory with a compact extra-dimension, in the $N_5\rightarrow \infty$ limit: it consists of a confining and of a dimensionally reduced phase, separated by a second-order phase transition for which the Polyakov loop along the compact extra dimension serves as the order parameter. The solid red curve, drawn to guide the eye, goes through the origin with diverging derivative, as expected from the strong-coupling expansion, and tends to an asymptotic value as $\widetilde{N}_5\rightarrow \infty$. The confining phase is analytically disconnected from the weak-coupling, large $\beta_5$ regime, and is an artifact of the lattice regularization.}
\label{fig:phase_diag}
\end{figure}

In order to map out the phase diagram, we measured the pseudo-critical coupling for fixed $\widetilde{N}_5$ and for various $N_5$ extrapolating to the $N_5 \rightarrow \infty$ limit, depicted in 
Fig.~\ref{finite_N5_effects_on_betac}\footnote{The discretization errors are of order $1/N_5$ as they arise from dimension six operators, some of which have coefficients proportional to $a a_5$.}. 
The resulting phase diagram in this limit is displayed in Fig.~\ref{fig:phase_diag}, where 
the two phases are separated by a second order transition line, instead of a first order transition as in the isotropic case. The order and universality class of the phase transition were established by a finite-size scaling study discussed in more detail
in the Appendix. At the transition, the correlation length of Polyakov loops in the fifth direction diverges, and, as expected from general arguments~\cite{Svetitsky:1982gs}, we found the transition to be in the trivial 4d Ising universality class. 

Finally, it is interesting to point out that some features of the $(4+1)d$ YM system are analogous to the more familiar case of an Abelian, U(1) gauge theory defined on a
$(3+1)d$ lattice, whose properties are well understood~\cite{Banks:1977cc, Frohlich:1982gf, Frohlich:1986sz}. 

It is well known that compact U(1) lattice gauge theory in 4d admits a dual formulation, see, for instance, Ref.~\cite{Panero:2005iu} and references therein, in terms of topologically non-trivial excitations which can be localized in elementary lattice cubes~\cite{DeGrand:1980eq} and interpreted as magnetic monopoles. Numerical investigations~\cite{Polley:1990tf} have revealed that such monopoles are responsible for the phase structure of the theory. In the weak-coupling regime ($\beta=1/e^2 \gg 1$) at zero temperature, the monopoles have a large mass $m(\beta)$, and their worldlines (which form closed loops, due to current conservation) are typically short, since the probability of a loop of size $L$ is proportional to $\exp[-m(\beta)L] \ll 1$. Such loops do not affect the long-distance physics: in the presence of external electric charges, the system exhibits Coulomb 
behavior (like its counterpart in the continuum limit, namely a photon gas), and the effect of the monopoles is simply to renormalize the electric charge. 

As $\beta$ is reduced and the bare coupling is made stronger, however, the monopoles become lighter, and their mass vanishes at a weakly first-order transition at $\beta_c = 1.0111331(21)$~\cite{Arnold:2002jk}---see also Ref.~\cite{Vettorazzo:2003fg}. At stronger coupling, monopoles condense and the system is in a confining phase, consistently with the strong coupling expansion of the lattice theory. Thus, the flux of the magnetic field coming from a monopole can be seen as disordering space-like Wilson loops. Therefore, at weak coupling, monopole loops of finite size can only induce a perimeter-like decay for the expectation values of asymptotically large Wilson loops. On the contrary, at strong coupling the monopole loops percolate, and disorder the Wilson loops on all scales, driving an exponential area-law decay and confining behavior.

Consider now the compact U(1) lattice gauge theory in a system with a compact fourth direction of size $L_4$: in the Coulomb phase at weak coupling, the probability of monopole loops winding around the compact direction is proportional to $\exp[-m(\beta) L_4]$. Such loops can disorder spatial Wilson loops on all length scales, and thus induce an area-law decay, because the magnetic flux is no longer neutralized by that of a neighboring anti-monopole. As a consequence, spatial Wilson loops will exhibit area law behavior, with a string tension proportional to the monopole loop density: $a^2 \sigma \propto \exp[-m(\beta) L_4]$. Such behavior has indeed been observed numerically in Ref.~\cite{Vettorazzo:2004cr}: the Coulomb phase gives way to a 3d confining phase for any finite $L_4$, where the global U(1) symmetry of the Polyakov loop is spontaneously broken. Moreover, by increasing $L_4$ while keeping $\beta$ fixed, one can increase the 3d correlation length exponentially: this achieves dimensional reduction and yields the continuum limit of the 3d Abelian theory, which is confining for all values of the coupling~\cite{Polyakov:1976fu, Gopfert:1981er}.

All this is completely analogous to the behavior of YM theory in $(4+1)$ dimensions in the weak-coupling phase, although here we have not been able to investigate the objects playing a role analogous to that of the magnetic monopoles as their gauge invariant formulation is yet to be found.

\section{Dimensional reduction}
\label{sec:dim_red}

In this Section, we discuss the properties of the dimensionally reduced phase of our model, comparing the perturbative predictions which are expected to hold in the weak (bare) coupling limit $\beta_5\rightarrow \infty$ and our numerical results from lattice simulations. 

Perturbatively, at large $\beta_5$ the theory has three mass scales, in analogy with the finite-temperature QCD setup in 4d~\cite{Ginsparg:1980ef, Appelquist:1981vg, Braaten:1995jr, Kajantie:2000iz}. Firstly, observe that the non-static Kaluza-Klein modes can be described by a tower of 4d modes with masses
\eq{Kaluza_Klein_mass_tower}
m_{KK}=\frac{2\pi}{L_5}n, \quad\quad n=1,2,\ldots .
\en
Secondly, at tree-level, the static Kaluza-Klein modes are all massless, but one-loop corrections give the gauge field in the fifth direction a mass proportional to
\eq{quantum_mass_for_static_modes}
m_5 \propto \frac{g_5}{L_5^{3/2}}.
\en
Finally, while the static modes of the gauge field components in the four ordinary directions remain massless to all perturbative orders they acquire a non-perturbative mass due to confinement in four dimensions. 

\begin{figure}  
\centerline{\includegraphics*[width=0.6\textwidth]{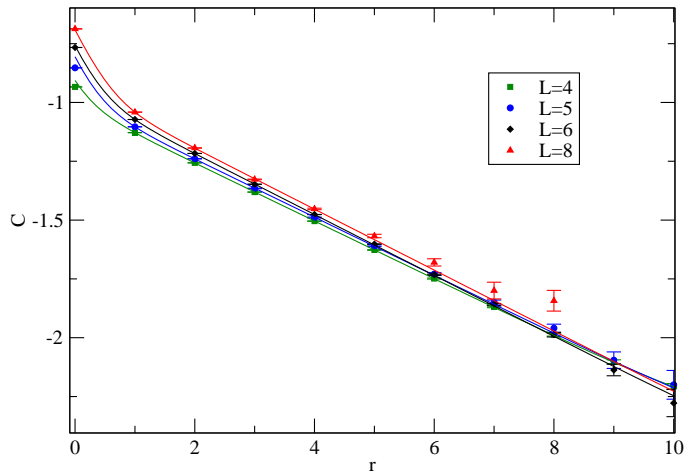}}
\caption{Two-point correlation function of zero-momentum torelons as a function of their separation, from lattices of sizes $6^2\times L \times 24 \times 2$ with $L=4,\dots,8$, $\beta=1.6575$, and $\gamma=1$. The four-dimensional L\"uscher correction has been subtracted from the correlators such that $C(r)=\log[\langle \tilde{P}_x(r) \tilde{P}_x(0)\rangle ]/L-\pi r/(3 L^2) $. The collapse of the curves at large distances indicates that the long distance properties of the model are described by a four-dimensional confining theory. At short distances the data also reveal the contribution from the first non-static Kaluza-Klein mode.}
\label{second_exponential}
\end{figure}

In the low-energy sector, namely at distances $\Delta x \gg L_5$, the system undergoes dimensional reduction, and the effective low-energy action describing the static Kaluza-Klein modes takes the form of a continuum 4d YM theory coupled to an adjoint scalar field $A_5$
\eq{EQCD_like_effective_action}
S_\rmi{eff} = \int {\dd}^4x  \left( \frac{1}{2}\tr F_{\mu\nu}^2 + \tr \left[ D_\mu A_5\right]^2 + m_5^2 \tr A_5^2+\lambda \tr A_5^4 \right),
\en
where the effects of the higher order terms are suppressed by powers of the scale ratio between the compactification scale ($L_5/2\pi$) and the typical scale of the effective theory $m_5^{-1}$. 

The 4d effective theory given by Eq.~(\ref{EQCD_like_effective_action}) is renormalizable, and is characterized by the parameters $g_4(\mu)$, $m_5^2(\mu)$ and $\lambda(\mu)$, where $\mu$ denotes the renormalization scale of the effective theory. At weak coupling, it is natural to impose a perturbative matching of the parameters and fields of the effective theory to the original theory, within the validity domain of the theory described by Eq.~(\ref{EQCD_like_effective_action}). Note that $A_M^{4d}$ and $g_4^2$ get their leading-order contribution at tree-level 
\eq{tree_level_contribution_to_AM_and_g4}
A_M^{4d}=\sqrt{L_5}A_M^{5d},\;\;\;\;\;g_4^2=g_5^2/L_5.
\en

\begin{figure} 
\centering
\includegraphics*[width=0.8\textwidth]{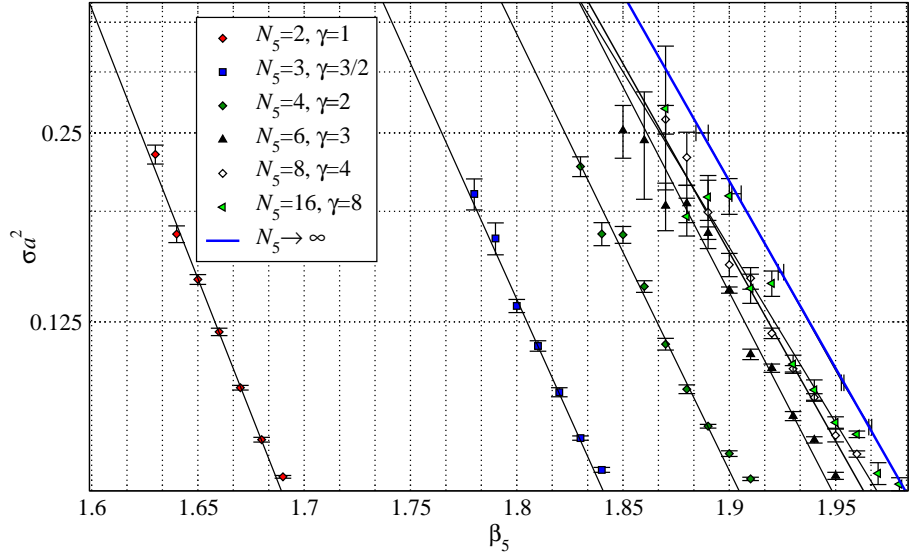}
\caption{A sample of our results for the string tension $\sigma a^2$, plotted as a function of $\beta_5$. The figure refers to data obtained from simulations at a fixed value of $\widetilde{N}_5=N_5/\gamma=2$, for which Eq.~(\ref{strng}) predicts a linear relation between $\ln (\sigma a^2)$ and $\beta_5$. The solid blue line on the right-hand side of the plot corresponds the results extrapolated to the $N_5 \to \infty$ limit, and its slope is consistent with the perturbative expectation.}
\label{fig:stringtension_vs_beta.L2}
\end{figure}

\begin{figure} 
\centering
\includegraphics*[width=0.5\textwidth]{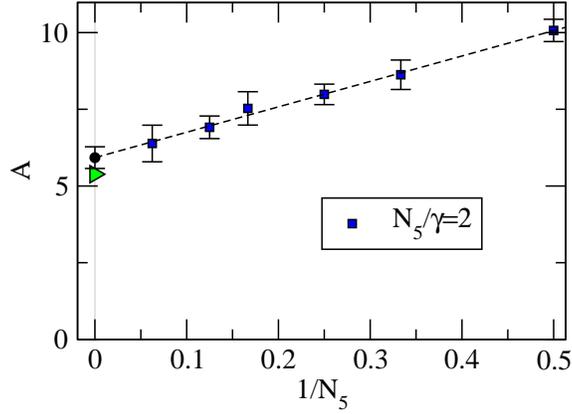}
\caption{ The coefficient $A$ describing the exponential decay of the string tension, as a function of the five-dimensional bare coupling $\sigma a^2 \sim \exp(-A \beta)$ as a function of $N_5$. The green triangle denotes the one-loop perturbative prediction as given by Eq.~(\ref{strng}). For lattices coarse in the fifth dimension, the coefficient differs by a factor of two, but approaching the continuum limit in the fifth dimension improves the agreement significantly.}
\label{fig:slope}
\end{figure}

On the other hand, the leading contributions to $m_5^2$ and $\lambda$ arise only at one-loop level, and can be readily evaluated by computing the effective potential using standard techniques. On an anisotropic Euclidean lattice in $d=5$ dimensions, the one-loop quantum fluctuations induce an effective potential for a classical background gauge field parametrized by $A_5(z)= \pi q(z) \sigma_3 / (L_5 g_5)$ of the form
\eq{1_loop_eff_pot}
V_{\mbox{\tiny{eff, 1-loop}}}^{(L)}(q) = -\frac{d-2}{L_5^d} \left( \frac{N_5}{\gamma} \right)^{d-1} \int_0^{\infty} \frac{ds}{s} \left[ e^{-2s} I_0(2s) \right]^{d-1} \sum_{n=0}^{N_5-1} e^{ - 4 s \gamma^2 \sin^2 \left[ \frac{\pi (n+q)}{N_5} \right] },
\en
from which the one-loop perturbative result for $m_5^2$ follows: 
\eq{one_loop_m5square}
m_5^2 = \left. \frac{1}{4} \left( \frac{\partial}{\partial A_5^3} \right)^2 V_{\mbox{\tiny{eff, 1-loop}}}^{(L)}(q)\right|_{q=0} .
\en

\begin{figure} 
\centering
\includegraphics*[width=0.8\textwidth]{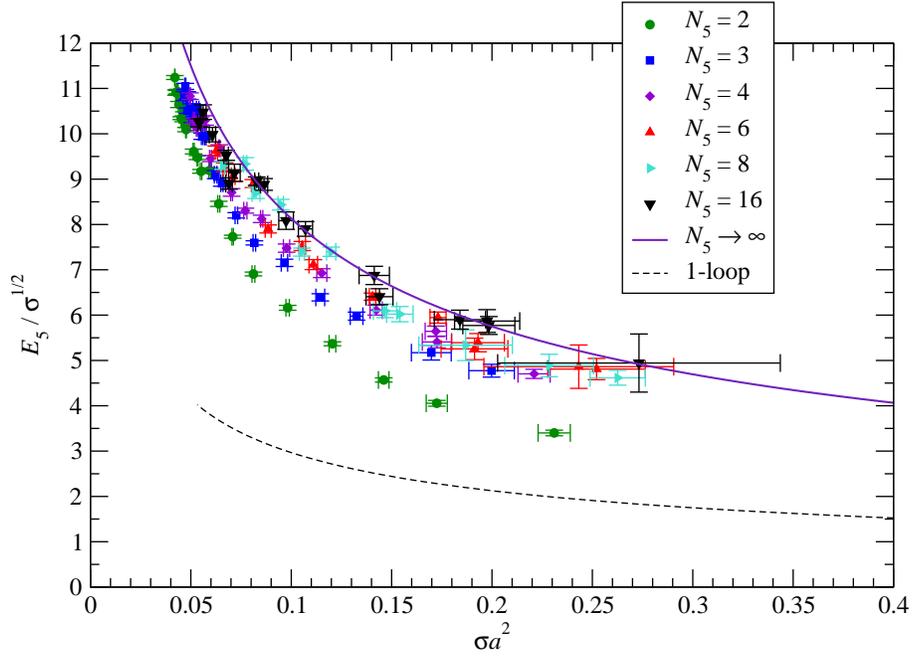}
\caption{Simulation results for the Higgs screening mass $E_5$ extracted from the exponential decay of the zero-momentum correlator of the Polyakov loop in the fifth direction, Eq.~(\ref{screening}), versus the string tension in units of the lattice spacing. The screening mass $E_5$ is expressed in units of the square root of the 4d string tension, and is given by $E_5=2 m_5$ to leading order in perturbation theory. The figure shows our data at fixed $\tilde N_5=N_5/\gamma=2$ and for different $N_5$ values. The solid curve is obtained by extrapolation to the $N_5 \to \infty$ limit. While the data show qualitative agreement with the one-loop perturbative prediction (thin dashed line), the quantitative details differ due to moderate values of the bare coupling $\beta$.}
\label{fig:a5mass_vs_stringtension2_L2}
\end{figure}

As the 4d components of the gauge field $A_\mu$ remain massless to all orders in perturbation theory, at length scales $\Delta x \gg m_5^{-1} $ the $A_5$ field can be integrated out completely, resulting in an effective theory which is just a 4d Yang-Mills model
\eq{MQCD_like_effective_action}
S'_{\rmi{eff}}=\int {\dd} ^4 x \left(\frac{1}{2}\tr F_{\mu\nu}^2\right),
\en
with a gauge coupling $g'_4(\mu)$.

As before, the coupling of the effective theory can be perturbatively matched to $g_4$ in the regime where both theories are expected to give a good description. Although the correlation length $\xi$ is infinite to all orders in perturbation theory, the 4d Yang-Mills theory described by Eq.~(\ref{MQCD_like_effective_action}) is known to be confining for all non-zero values of $g'_4$, since it non-perturbatively develops a mass gap, and thus a finite correlation length $\xi = 1/\sqrt{\sigma}$ inversely proportional to the square root of the associated 4d string tension $\sigma$. At leading order, asymptotic freedom requires the dependence of the correlation length on the renormalized gauge coupling $g'_4(\mu)$ to be of the form
\eq{xi}
\xi=\frac{c}{\mu}\exp\left[\frac{1}{2 b_0 g'^2_4(\mu)}\right],
\en
where $b_0=11/(24 \pi^2)$ is the coefficient of the leading term in the $\beta$-function, and $c$ is a constant to be determined non-perturbatively. This is the origin of Eq.~(\ref{mass_gap}). Correspondingly, the large-distance decay of Wilson loops normal to the compactified direction is expected to be characterized by a string tension
\eq{strng}
\sigma_{4d}  \sim \frac{1}{L_5^2} \exp \left[ - \frac{1}{b_0 g'^2_4(L_5)}\right] \sim \frac{1}{a^2 \widetilde{N}_5^2} \exp \left[ - \frac{\beta_5\widetilde{N}_5}{4b_0} \right]. 
\en

Our simulations show that the scale separation and the dimensional reduction carry over also to non-zero couplings. 
In Fig.~\ref{second_exponential}, we plot zero transverse momentum torelon correlators for various torelon lengths $L$ as a function of separation. The four-dimensional L\"uscher-term has been subtracted from the correlators, such that the collapse of the curves at large separations (absent if one applies a five-dimensional correction) indicates that the long distance properties of the five-dimensional theory are described by a four-dimensional confining theory, as anticipated by the weak coupling analysis. While at large distances the correlator is well described the four-dimensional theory, at short distances hints of the first non-static Kaluza-Klein mode can be seen. Indeed, a two-exponential fit to the data, with the second mass being that appropriate for the first non-static mode (given by Eq.~(\ref{Kaluza_Klein_masses})) captures the short distance behavior; the extrapolation of the fitting function all the way to zero distance reproduces the data reasonably well.

Having established the confining nature of the long-distance sector we move on to test the effective theory prediction of the behavior of the string tension as a function of the parameters of the five-dimensional theory. In Fig.~\ref{fig:stringtension_vs_beta.L2} we display a collection of string tensions measured from the torelon correlators for several values of $N_5$, keeping $\widetilde{N}_5$ fixed and varying the bare coupling $\beta$. For all values of $N_5$, the data points form a straight line in a semi-logarithmic plot
\ba
\sigma a^2 \sim \exp(-A \beta),
\ea
as anticipated by Eq.~(\ref{strng}). However, for small values of $N_5$ the slope, $A$, is significantly larger than expected from the perturbative prediction. Nevertheless, taking the continuum limit in the fifth direction brings us significantly closer to the one-loop result as depicted in Fig.~\ref{fig:slope}. While the bare coupling $\beta$ was not that small in the simulations ($\sim 1.6-2$) the good agreement between perturbation theory and simulations is perhaps not completely unexpected as the coefficient in the exponential of Eq.~(\ref{strng}) is protected from a two-loop correction. The evaluation of the $g'^2_4(L_5)$ to two loops would only induce an overall shift in the amplitude $c$ and thus the first non-trivial correction enters at the three-loop level.

We also measured the correlators of Polyakov loops in the fifth direction from the same data set, whose exponential decay
\ba
\label{screening}
\langle \tilde{P}^\dagger_5(r) \tilde{P}_5(0) \rangle\sim \exp(-E_5 r),
\ea
is given to leading order in perturbation theory by $E_5=2m_5$. The results in units of string tension are shown Fig.~\ref{fig:a5mass_vs_stringtension2_L2}, where it can be seen that the
measured data are qualitatively described by the leading order perturbation theory, but with sizable quantitative differences.

\section{Towards the continuum}
\label{sec:contlim}
\begin{figure}  
\centering
\includegraphics*[width=12cm]{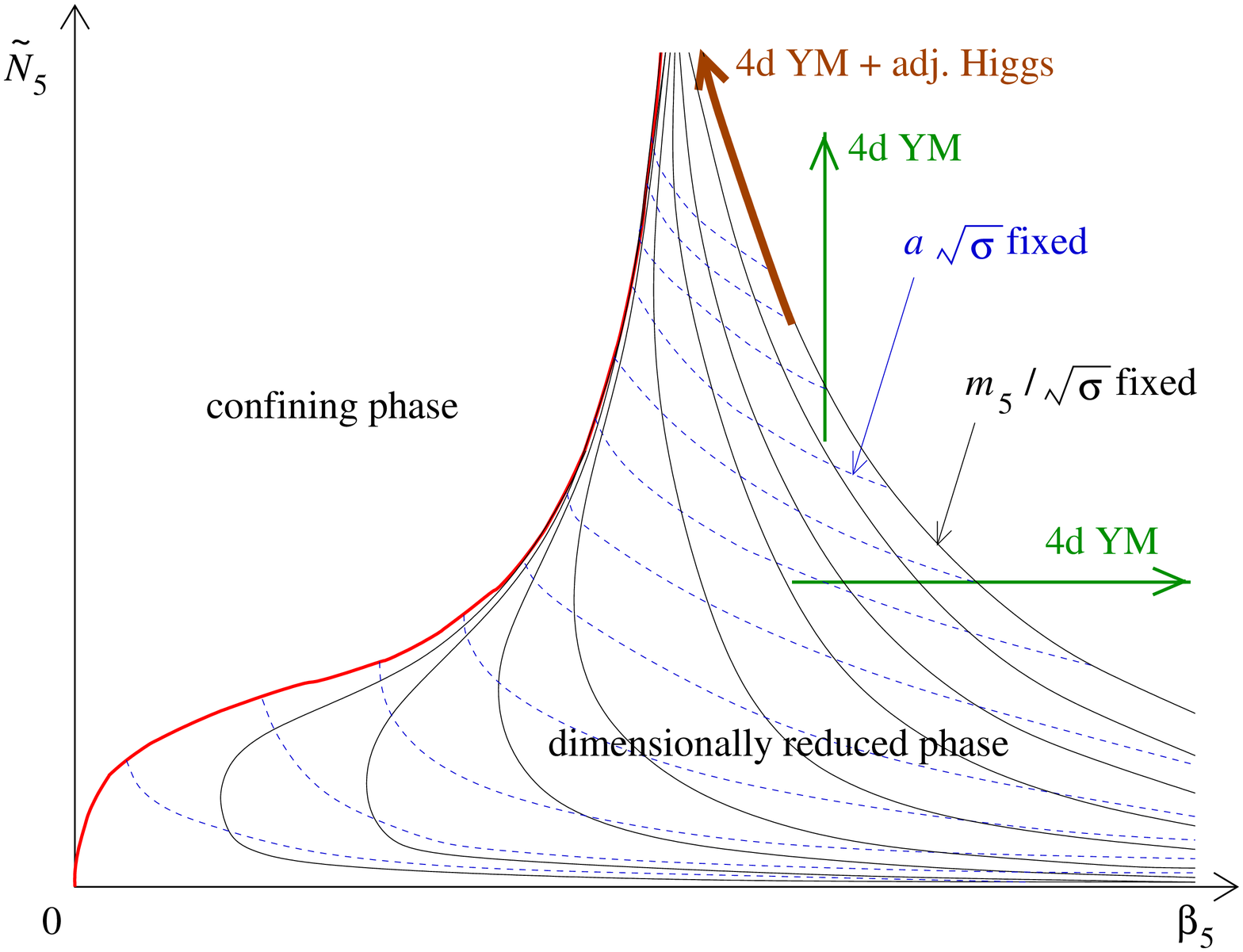}
\caption{Sketch of the lines of constant low-energy physics (labelled by the $m_{5}/\sqrt{\sigma}$ ratio) in the $(\beta_5,\tilde N_5)$ phase diagram Fig.~\ref{fig:phase_diag} of the 5d lattice model after taking the continuum limit in the fifth direction only. The red, leftmost, line indicates the phase transition separating the confining and the dimensionally reduced phases of the lattice theory. The lines of constant physics are displayed as solid black lines; at large $\beta_5$ values, they tend to hyperboles of equation $\beta_5 \tilde N_5 = \mbox{const}$. The dashed blue lines correspond to fixed values of $\sigma a^2$: the lattice spacing $a$ decreases when $\beta_5 \to \infty$ at fixed $\tilde N_5$ (green horizontal arrow), or when $\tilde N_5 \to \infty$ at fixed $\beta_5$ (green vertical arrow). In both these limits the $A_5$ field decouples, and one recovers the ordinary continuum limit of the 4d YM theory. On the contrary, a continuum limit including 4d YM coupled to an adjoint scalar Higgs can be obtained by moving upwards along a line of fixed $m_{5}/\sqrt{\sigma}$ (brown arrow). Finally, observe that, for any fixed, finite value of $\sigma a^2$ (i.e. on any dashed blue line), there is a finite maximum value of $\tilde N_5$, i.e. a finite maximum value of $L_5/a$, which is reached at the $\beta_5$ value where the dashed blue line hits the phase transition line.}
\label{fig:LCP_sketch_with_a}
\end{figure}

Let us now discuss the lines of ``constant low-energy physics'' in the dimensionally-reduced phase of the lattice model, meaning the lines corresponding to a fixed value of the ratio $m_5/\sqrt{\sigma}$ of the $A_5$ field mass over the typical energy scale $\sqrt{\sigma}$ characterizing the 4d physics. In particular, in the following we discuss such lines in the phase diagram obtained after taking the continuum limit in the fifth direction, $N_5 \rightarrow \infty$, as a function of $\beta_5$ and $\widetilde N_5$, as sketched in Fig.~\ref{fig:LCP_sketch_with_a}.

Starting from very large $\beta_5$ values, leading-order perturbation theory predicts that the lines of fixed $m_{5}/\sqrt{\sigma}$ tend asymptotically to hyperboles of equation $\beta_5 \widetilde N_5 = \mbox{const}$ (in particular, $m_{5}/\sqrt{\sigma}$ increases when $\beta_5 \widetilde N_5$ takes larger and larger values). As one moves along one such line towards smaller values of  $\beta_5$, the ultraviolet cutoff measured in units of the 4d string tension, $\sigma a^2 $,
decreases monotonically, and corrections which depend on $a/g_5^2$ make the line deviate from its asymptotic hyperbolic behavior. Following the line further upwards in the $(\beta_5,\widetilde N_5)$ plane, it eventually approaches the phase transition line, and bends to follow it---without crossing it, as long as the latter is of second order (as this would imply $m_5/\sqrt{\sigma}=0$). 

In Fig.~\ref{fig:LCP_sketch_with_a} we also display the pattern of lines parametrized by $\sigma a^2=\mbox{const}$, i.e. by a fixed value of the 4d lattice spacing, as dashed blue lines. The value of $a$ decreases as one moves upwards along a vertical straight line, or to the right along a horizontal line. 

Moving along a line of fixed $m_{5}/\sqrt{\sigma}$ from very large to smaller $\beta_5$ values, the lattice spacing $a$ decreases. This is not in contradiction with the intuitive expectation that the lattice spacing should become smaller when $\beta_5$ is increased towards $+\infty$, if $\widetilde N_5$ is held fixed. By taking the $\beta_5 \to \infty$ limit at fixed $\widetilde N_5$ (which is described by the green horizontal arrow in Fig.~\ref{fig:LCP_sketch_with_a}), one crosses lines corresponding to larger and larger $m_{5}/\sqrt{\sigma}$ values. Note that this corresponds to moving along the extrapolated violet curve appearing in Fig.~\ref{fig:a5mass_vs_stringtension2_L2}, going towards the vertical axis $\sigma a^2=0$. In such a limit, the $A_5$ static modes (and the Kaluza-Klein tower) become infinitely heavy compared to the 4d energy scale, and decouple from the low-energy sector of the theory. Thus, in this limit the model simply reduces to the ordinary continuum 4d YM theory, which is renormalizable, and for which there is no obstruction in removing the ultraviolet cutoff $a \rightarrow 0$.  

More generally, this ``ordinary'' 4d continuum limit can be reached by considering any trajectory that crosses lines corresponding to increasing values of the $m_5/\sqrt{\sigma}$ ratio, while increasing the 4d correlation length $\xi$ in units of $a$. As an example, the $A_5$ field completely decouples from the 4d physics also when one moves along the vertical line in Fig.~\ref{fig:LCP_sketch_with_a} (at a fixed $\beta_5$ larger than the limiting value of the critical coupling which is reached for $\widetilde N_5 \to \infty$), as was suggested in Refs.~\cite{Chandrasekharan:1996ih, Beard:1997ic, Brower:2003vy}. If one interprets the 5d model as an effective theory of
a more fundamental theory defined with a cutoff $a$, this feature has a striking consequence: by ever increasing the extent of the extra-dimension in units of the cutoff, one ends up with a four-dimensional theory with no trace of its extra-dimensional ancestry.

On the other hand, the cutoff can also be decoupled by approaching the phase transition line moving along a line of fixed $m_5/\sqrt{\sigma}$ (and therefore of fixed $m_5$, if one assumes that the 4d physics scale $\sigma$ is fixed): this is the case in which the resulting continuum theory features a scalar field coupled to 4d YM. To remove the cutoff in the 4d model at fixed $m_5/\sqrt{\sigma}$ one should study the lattice model at points along a line like, e.g., the brown curve in Fig.~\ref{fig:LCP_sketch_with_a}. In this limit, going to larger and larger $\tilde N_5$ values means that $\sigma a^2$ tends to zero (continuum limit) and also $am_5$ goes to zero due to the approach of the second order transition line. However, the non-static Kaluza-Klein modes decouple in this limit, as it is only the static Kaluza-Klein mode which becomes critical and whose correlation length diverges at the second order transition.

In short, our 5d lattice YM model with one compactified dimension  admits two different types of continuum limits in its dimensionally reduced phase:
\begin{itemize}
\item ordinary 4d YM theory;
\item 4d YM theory coupled with an adjoint scalar field.
\end{itemize}
Both limits are well-defined, renormalizable four-dimensional quantum field theories.

\section{Conclusions}
\label{sec:conclusions}
In this article, we have presented a non-perturbative lattice study of 5d SU(2) YM theory with one compact extra dimension. This model is a simple example in a class of theories which could be interesting for TeV physics and physics beyond the Standard Model, if large extra dimensions are realized in nature and can be probed by the forthcoming high-energy experiments at the Large Hadron Collider. 

At weak (bare) coupling and for all compactification lengths $L_5/a$, the model exhibits a pronounced scale separation, which grows exponentially as a function of $L_5/a$, between the static KK-modes of the four-dimensional gauge fields and all the other modes. 
In this paper we have presented numerical evidence showing that this picture of dimensional reduction  carries over to non-zero couplings as well. 

As the correlation length of the lightest modes is inevitably very large compared to the cutoff scale, the 
effects of the details of the regularization are strongly suppressed for low-energy observables. In particular,
the rotational symmetry is approximately restored. This defines a sensible scheme to obtain non-perturbative information about the low-energy sector from simulations of the five-dimensional lattice theory.

In particular, using this approach it is possible to derive an upper bound for the radius of the extra-dimension for a given four-dimensional correlation length and ultra-violet cutoff (see Fig.~\ref{fig:LCP_sketch_with_a}: when $\sigma a^2$ is fixed, $L_5/a \approx \tilde N_5$ is bounded). For example, taking the electroweak scale as the 4d scale and assuming the cutoff to be at the Planck scale gives a maximum compactification length
of the order of $10M_P^{-1}$. Also, the large scale difference can potentially be used to acquire information
about the compactification and cutoff scales from experimental measurements of four-dimensional quantities. If the mass $m_5$ and the self-coupling $\lambda$ of a scalar Higgs-like particle can be determined, say, at future LHC experiments, this information can be non-perturbatively mapped via lattice simulations to the two parameters of our lattice model $\beta_5$,$\widetilde{N}_5$, from which both the compactification scale and the cutoff scale can be estimated.
For example, Fig.~\ref{fig:a5mass_vs_stringtension2_L2} illustrates how an
experimental measurement of $m_5/\sigma^{1/2}$ can be used as input to
determine $\sigma a^2$, and thereby the cutoff $a^{-1}$.

Our main results are summarized in Fig.~\ref{fig:phase_diag} and in Fig.~\ref{fig:LCP_sketch_with_a}. Fig.~\ref{fig:phase_diag} shows the general structure of the phase diagram of the theory after extrapolation to the continuum limit \emph{only in the fifth direction}, which is achieved by letting $N_5 \rightarrow \infty$ (while keeping the 4d lattice spacing $a$ finite), as a function of $\beta_5=4a/g_5^2$ and $\tilde N_5 = N_5/\gamma$. Fig.~\ref{fig:LCP_sketch_with_a} displays the behavior of the lines of constant low-energy physics (solid black curves) parametrized by constant values of the $m_5/\sqrt{\sigma}$ ratio ($m_5$ being the non-vanishing mass associated with the static-modes of the $A_5$ field) in the dimensionally reduced phase. 

The cutoff can be sent to infinity in two different ways, both of which correspond to well-defined, renormalizable quantum field theories: 4d YM theory, and 4d YM theory coupled to an adjoint scalar field. The numerical results of our simulations are consistent with those of a similar study~\cite{Ejiri:2000fc,Ejiri:2002ww}, however our approach (based on a dynamical definition of the lattice spacing $a$ from the four-dimensional string tension) leads to different physical conclusions. In particular, we emphasize that our findings are completely consistent with the non-renormalizability of the 5d YM theory. One can construct continuum limits yielding ordinary four-dimensional models, while it is impossible to define a full-fledged continuum limit preserving the 5d nature of the YM theory, including the whole tower of Kaluza-Klein modes\footnote{We have found no indication in our study
for a non-trivial UV fixed point in a 5d isotropic system, as conjectured in \cite{Peskin:1980ay,Morris:2004mg,Gies:2003ic} and investigated on the lattice in \cite{Kawai:1992um}}.

In summary, the present study has clarified the phase structure of 5d YM theory, discussing the features of the phase diagram obtained from lattice simulations in a regime where a separation of scales makes the low-energy physics insensitive to the details of the theory at the cutoff scale. 

As for future lines of research, possible extensions of this study include:
\begin{itemize}
\item extension to larger non-Abelian gauge groups; although it is reasonable to expect that the general qualitative features of 5d lattice YM theories are already captured by the present study of an SU(2) model, the extension to higher-rank groups might lead to a variety of spontaneous symmetry breaking patterns;
\item inclusion of fermionic degrees of freedom; in particular, a 5d model could provide a natural setup to study electroweak symmetry breaking and related problems. Furthermore, it would also be interesting to discuss the generalization of the setup studied in the present work by including lattice fermions in the domain-wall formulation~\cite{Kaplan:1992bt, Furman:1994ky}.
\end{itemize}

\section*{Acknowledgements}

PdF thanks KITPC for hospitality. AK acknowledges support from the SNF grant 20-122117. MP acknowledges support from INFN. We warmly thank Konstantinos Anagnostopoulos, Konstantinos Farakos, Leonardo Giusti, Maarten Golterman, Keijo Kajantie, Tony Kennedy, Mikko Laine, Stam Nicolis, Kari Rummukainen, Claudio Scrucca, Antonios Tsapalis and Uwe-Jens Wiese for enlightening discussions. The numerical simulations were performed at the Center for Scientific Computing (CSC), Finland under the HPC-Europa2 project (project number: 228398) with the support of the European Commission - Capacities Area - Research Infrastructures.

\section*{Appendix: Finite-size scaling analysis of the phase transition}
\renewcommand{\theequation}{A.\arabic{equation}}
\setcounter{equation}{0}

To ascertain the order of the phase transition, we performed a finite-size scaling analysis for a $N^4\times2$ lattice at $\gamma=2$ for a range of volumes, measuring the fourth-order Binder cumulant associated with the average Polyakov loop along the fifth direction:
\eq{Binder_cumulant_definition}
B_4=1-\frac{\langle P_5^4 \rangle}{3\langle P_5^2 \rangle^2}.
\en

\begin{figure}  
  \centering
\includegraphics*[width=0.8\textwidth]{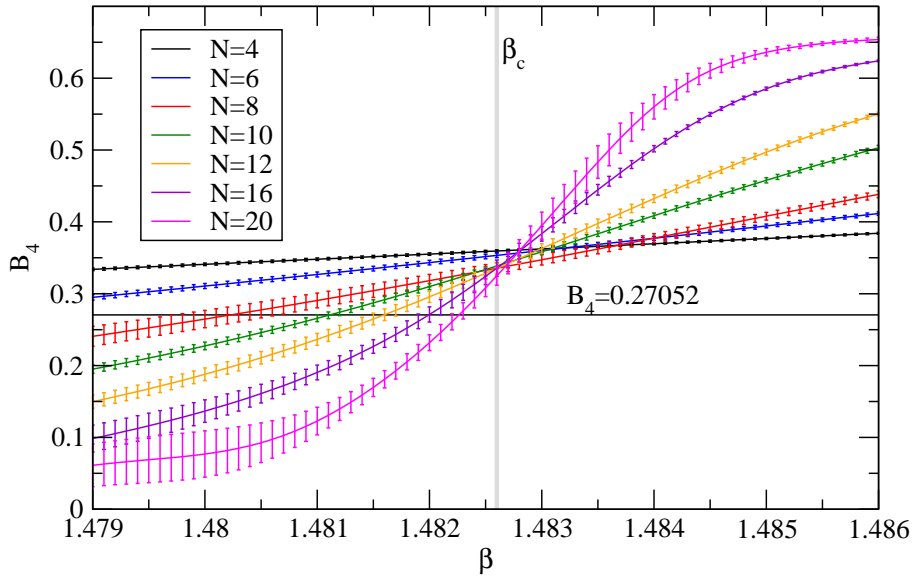}
\caption{Finite-size scaling of the fourth-order Binder cumulant $B_4=1-\langle P_5^4 \rangle/(3\langle P_5^2 \rangle^2)$ on a $N^4\times2$ lattice with $\gamma=2$. The horizontal line shows the mean-field value for the 4d Ising model and the vertical line indicates our determination of the critical value of $\beta$.}
\label{fig:Binder}
\end{figure}

The Binder cumulants measured on lattices with various volumes (shown in Fig.~\ref{fig:Binder}) as a function of $\beta$ are expected to cross at the critical point, at a universal value of $B_4$ characteristic of the universality class, provided that the transition is of second order. The transition breaks the $Z_2$ symmetry of the Polyakov loop and thus the universality class should be that of the 4d Ising model. 

Since four is the upper critical dimension for the Ising model, the critical value of the Binder cumulant can be estimated by considering a mean-field effective action with only one degree of freedom~\cite{Parisi:1996vu}:
\ba
Z_{\text{eff}}&=&\int \dd \phi~\exp(-S_\text{eff})\\
S_{\text{eff}}&=&N^4\left(\frac{1}{2}t \phi^2+\frac{1}{4!}u \phi^4\right),
\label{eq:S_4dIsing}
\ea
where $t$ is the reduced temperature and $\phi$ the mean field. 

Note that there is no contradiction between the effective action Eq.~(\ref{eq:S_4dIsing}) where $u\neq 0$, and the triviality of $\phi^4$ theory in four dimensions. The latter states that the \emph{renormalized} coupling constant vanishes, whereas $\phi$ in Eq.~(\ref{eq:S_4dIsing}) stands for the \emph{bare} field measured in the simulation: the mean $P_5$ Polyakov loop. 

\begin{figure}
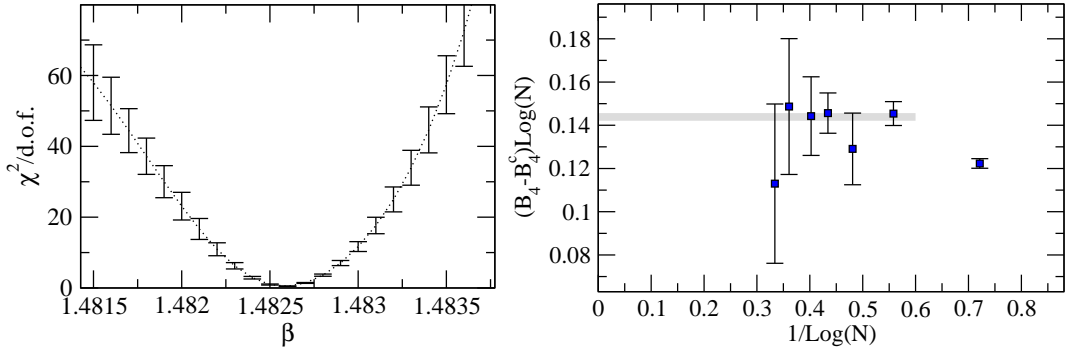
  
  \centering
\includegraphics*[width=6.55cm]{fig9a.eps}
\includegraphics*[width=7.45cm]{fig9b.eps}
\caption{On the left-hand panel, the reduced $\chi^2$-value of the fitting ansatz Eq.~(\protect\ref{behavior_at_upper_critical_dimension}), applied to the data of Fig.~\ref{fig:Binder} (the points corresponding to $N=4$ were not included in the fit). On the right-hand panel, the data at $\beta=1.4826$ are well described by the ansatz Eq.~(\protect\ref{behavior_at_upper_critical_dimension}), with $\chi^2/\text{d.o.f}=0.50(7)$ and quickly deteriorates away from this point, which we take as our estimate of the critical point.}
\label{fig:chi2}
\end{figure}

At criticality, i.e. for $t=0$, one gets:
\eq{Binder_cumulant_critical_value}
B_4^c=1-\frac{\Gamma^2(1/4)}{12 \Gamma^2(3/4)} \approx 0.27052 .
\en
Furthermore, the fact that four is the upper critical dimension for the Ising model also implies that the corrections to the critical value of $B_4$ decrease only logarithmically with the linear extent of the system:
\eq{behavior_at_upper_critical_dimension}
B_4(N)=B_4^c+A/\ln(N). 
\en
While rendering the direct observation of the crossings of the Binder cumulants numerically unfeasible, this form allows one to estimate a second-order critical point as the value of $\beta$ where the $N$-dependence of the Binder cumulant is well described by the ansatz. Fig.~\ref{fig:chi2} shows the reduced $\chi^2$ values from fits to Eq.~(\ref{behavior_at_upper_critical_dimension}) at various constant couplings $\beta$, and the fit at the critical coupling. We conclude that our data are compatible with a second order phase transition in the universality class of the 4d Ising model at $\beta_c=1.4826(1)$. 

\end{document}